\documentstyle [12pt] {article}

\begin{document}
\begin {center}
{\Large \bf Hamiltonian Systems on Quantum Spaces}\\
\vspace{1cm}
{\bf A. Shafei Deh Abad}\\
\vspace{7mm}

{\it Institute for Studies in Theoretical Physics and Mathematics(IPM)\\
P.O.Box 19395-5746\\
Tehran, Iran}\footnote{mailing address\\
e-mail: shafei@Irearn.Bitnet and shafei@netware2.ipm.ac.ir}\\
{\it Department of Mathematics, Faculty of Sciences, University of Tehran\\
Tehran, Iran}\footnote{permanent address}\\
\end{center}
\vspace{3mm}

\begin{abstract}
In this paper we consider Hamiltonian systems on the quantum plane and we
show that the set of Q-meromorphic Hamiltonians is a Virasoro algebra with
central charge zero and the set of Hamiltonian derivations of the algebra of
$Q$-analytic functions ${\cal A}_q$ with values in the algebra of
$Q$-meromorphic functions ${\cal M}_q$ is the Lie algebra $sl(2,A_1(q)).$
Moreover we will show that any motion on a quantum space is associated with
a quadratic Hamiltonian.
\end{abstract}
\newpage
{\bf Introduction}

Classical and quantum mechanics on q-deformed spaces have been studied by
many authors. Most of these works are concerned Hamiltonian systems, but
there are also some works about Lagrangian formalism on the quantum plane.
In all these works the q-deformed symplectic structure is obtained by the
q-deformation of the natural symplectic structure of the plane and it enables
one to obtain the equations of motion in the form
$dx/dt=\{H,x\}_q$,$dp/dt=\{H,p\}_q$. Unfortunately the q-deformed
Poisson bracket has nothing in common with the usual Poisson bracket and
the only use of it is in writting the equations of motion as above.
But most of the very interesting facts of classical mechanics are absent here.
It is unfotunate that here, in general
it is not true that $\{H,H\}_q=0$, and $\{H,f\}_q=0$ does not imply that
$\{H^2,f\}_q=0$.
In this situation what can be said about the Liouville's integrability
theorem?

The interpretation of the quantum spaces given in [3] enables us to have
Newtonian mechanics on these spaces. But it is exactly the same as Newtonian
mechanics on the ordinary affine spaces.On the other hand, it is well-known
that there are two other approaches to classical mechanics based on the
symplectic structure of the phase space [1].
The first is the state approach and the second is the observable approach.
In these approaches the coordinate and the momentum functions appear like
other observables and they all satisfy the same equation. On quantum spaces,
the state approach leads to the fact that the mass is a c-number.
While accoording to the observable approach the mass is an operator.
Therefore, on quantum spaces these two approaches are not equivalent.
Moreover, they are different from Newtonian approach to the classical
mecanics on quantum spaces.
In this paper following [5,6] we follow the observable approach.
Hence the mass will be considered as an operator.
To be more precise, let ${\cal M}_q$
denote the $A_0(q)$-algebra of q-meromorphic functions and $\pi$ be the
canonical
q-deformed Poisson structure on ${\cal M}_q$. By a Hamiltonian system we
mean a triple
$({\cal M}_q,\pi,z)$, where $z$ is a $\pi$-Hamiltonian element of
${\cal M}_q$. Here by a motion of the above system we mean a one-parameter
group of automorphisms of the system ${\phi}_t$, satisfying the following
condition
$$\forall f \in {\cal M}_q
\;\;,\;\;{\frac{{d{\phi}_t}(f)}{dt}}=\{z,{{\phi}_t}(f)\}.$$
Notice that we did not assume from the begining that the mass is a constant
of motion, but it will be proved.
In this way we see that we can not consider an arbitrary element of
${\cal M}_q$ as a Hamiltonian.
Indeed, the set of all Hamiltonians constitute a Virasoro algebra with
central charge zero. But as we will see
for a general Hamiltonian in our sense the corresponding Hamilton equations,
in general does not
define any motion. But when we restrict ourselves to the Hamiltonian systems
of the form
$({\cal A}_q,\pi,z)$, we see that the Hamilton equations define the motion of
the
system. These motions
are of very restricted types. Generally speaking, using Proposition 1.1 one
can easily see that the only possible motions on
the quantum spaces
are those associated with quadratic Hamiltonians. This fact suggests that we
should look for other quantum manifolds to have motions of other types.
We emphasize that it is easy to see that
the state approach also gives the same result.

Before going further we remind that in this paper our notations and
conventions are as in [3]. Moreover,
here by $A_0(q)$ we mean the C-algebra of all absolutely convergent power
series $\sum_{i>-\infty} {c_i{q^i}}$ on
$D-{\{0\}}$ with values in C, and by ${\cal A}_q$ we mean
the $A_1(q)$-algebra of Q-analitic functions on the q-deformed
R$^3$ with the following commutation relations between the coordinate
functions $x,p,m,$
$$x^ax^b=x^{a+b}, p^ap^b=p^{a+b}, m^am^b=m^{a+b}, p^ax^b=q^{ab}x^bp^a,
m^ax^b=q^{ab}x^bm^a, m^ap^b=q^{ab}p^bm^a,$$
where $a$ and $b$ are in Z. Also, by ${\cal M}_q$ we mean
the $A_0(q)$-algebra
of Q-meromorphic functions of the form
$$\sum_{i,j,k \gg-\infty} a_{ijk}(q)x^ip^jm^k,$$
where the sign $"\gg"$ under the
$"\Sigma"$ means that the indices $i,j,k$
are bounded below. We say that the function $z$ is a generalized
Q-meromorphic function if it is of the following form
$$z=\sum_{{i,j,k}>-\infty}{a_{ijk}}(q)x^ip^jm^k.$$
Notice that the set of generalized Q-meromorphic functions does not
constitute an algebra.
Finally, throughout the paper the sign $"-"$ on a $"\Sigma"$ means that
the $"\Sigma"$
is with finite support.

{\bf 1 Derivations of ${\cal M}_q$}

A linear operator $D:{\cal M}_q \rightarrow {\cal M}_q$ is called a
derivation if
for each $z_1,z_2\in {\cal M}_q$
$$D(z_1z_2)=D(z_1)z_2+z_1D(z_2).$$
{\bf Proposition 1.1.} A linear operator $D:{\cal M}_q \rightarrow
{\cal M}_q$ is a derivation
if and only if

$1) \hspace{7.5mm} qAp=pA \hspace{7.5mm}, \hspace{7.5mm} Bx=qxB$

$2) \hspace{7.5mm} qBm=mB \hspace{7.5mm}, \hspace{7.5mm} Cp=qpC$

$3) \hspace{7.5mm} qAm=mA \hspace{7.5mm}, \hspace{7.5mm} Cx=qxC,$

where $A=D(x)\;\;,\;\;B=D(p)\;\;,\;\;C=D(m).$

{\bf Proof.} Assume that $D:{\cal M}_q\rightarrow {\cal M}_q$ is a derivation
and let
$$A=\sum_{i,j,k\gg-\infty} a_{ijk}(q)x^ip^jm^k ,
B=\sum_{i,j,k\gg -\infty} b_{ijk}(q)x^ip^jkm^k.$$
Then, $$qD(xp)=qD(x)p+qxD(p)= $$
$$\sum_{i,j,k\gg-\infty} [a_{ijk}(q)q^{k+1}x^ip^{j+1}m^k+
b_{i,j,k}(q)qx^{i+1}p^jm^k]=D(px)=$$
$$Bx+pA=\sum_{i,j,k\gg-\infty} [a_{ijk}(q)q^ix^ip^{j+1}m^k+
b_{ijk}(q)q^{k+j}x^{i+1}p^jm^k].$$
Therefore for each i,j,k we have
$$ \;\;\;\;(*) \hspace{2cm}a_{i(j-1)k}q^{k+1}+b_{(i-1)jk}
q=a_{i(j-1)k}q^i+b_{(i-1)jk}q^{k+j}.$$
Clearly, if $b_{(i-1)jk}=0$ then $i=k+1$ and if $a_{i(j-1)k}=0$ then $k+j=1$.
Now, assume that $k+1=i$.
Then, $a_{(i-1)jk}\neq 0$ implies that $k+j=1$.
Assume that $k+j=1$. Then $a_{i(j-1)k}\neq 0$ implies that
$k+1=i.$ Now suppose that $a_{i(j-1)k}\neq 0 \neq b_{(i-1)jk}$ and $k+j=1$.
Then from $(*)$ we have
$$\;\;\;\;(**) \hspace{2cm}
{\frac{q^{k+1}-q^i}{q^{k+j}-q}}a_{i(j-1)k}=b_{(i-1)jk}.$$
It is clear that without any loss of generality we may assume that
$$D(x)=x^ap^{b-1}m^c ,   D(p)={\frac{q^{c+1}-q^a}{q^{b+c}-q}}x^{a-1}p^bm^c.$$
Then direct computation shows that
$$D(x^mp^n)=D(x^m)p^n+x^mD(p^n)=$$
$$(q^{b+c}-q)^{-1}[{\sum_{i+j=m-1} q^{(j+1)(b+c)+nc-j}}-{\sum_{i+j=m-1}
q^{j(b+c-1)+nc+1}}-$$
$${\sum_{i+j=n-1}q^{(j+1)c+i(a-1)+1}}
+\sum_{i+j=n-1}q^{jc+a(i+1)-i}x^{m+a-1}p^{n+b-1}m^c].$$
Therefore, from
$$D(x^mp^nx^rp^s)=q^{nr}D(x^{m+r}p^{n+s})=x^mp^nD(xrp^s)+D(x^mp^n)x^rp^s,$$
we have
$${\sum_{i+j=m+r-1}q^{nr+(j+1)(b+c)+(n+s)c-j}}-
{\sum_{i+j=m+r-1}q^{nr+j(b+c-1)+(n+s)c+1}}$$
$$-{\sum_{i+j=n+s-1}q^{nr+(j+1)c+i(a-1)+1}}+
{\sum_{i+j=n+s-1}q^{jc+a(i+1)-i+nr}}$$
$$-{\sum_{i+j=r-1}q^{n(r+a-1)+(j+1)(b+c)+sc-j}}+
{\sum_{i+j=r-1}q^{n(r+a-1)+j(b+c-1)+sc+1}}$$
$$+{\sum_{i+j=s-1}q^{n(r+a-1)+(j+1)c+i(a-1)+1}}-
{\sum_{i+j=s-1}q^{n(r+a-1)+ic+a(i+1)-i}}$$
$$-{\sum_{i+j=m-1}q^{c(r+s)+r(n+b-1)+(j+1)(b+c)+nc-j}}+
{\sum_{i+j=m-1}q^{c(r+s)+r(n+b-1)+j(b+c-1)+nc+1}}$$
$$+{\sum_{i+j=n-1}q^{c(r+s)+r(n+b-1)+(j+1)c+i(a-1)+1}}-
{\sum_{i+j=n-1}q^{c(r+s)+r(n+b-1)+jc+a(i+1)-i}}=0.$$
But since $b+c\neq 1$, the coefficient of the term
$q^{n(a-1)+r(n+b+c-1)+sc-(b+c)+2}$ appearing in the sixth $"\Sigma"$
is not zero in the above equation.
This contradiction proves that the last case isnot possible.
Therefore, $pA=qAp$ and $Bx=qxB.$
The proofs of $2)$ and $3)$ are the same.

Convesely, let $D:{\cal M}_q\rightarrow {\cal M}_q$ be a linear operator
and $A=D(x)$,  $B=D(p)$ and $C=D(m)$. Assume that
$$pA=qAp\,\,,\,\,mA=qAm\,\,,\,\,Bx=qxB\,\,,\,\,
mB=qBm\,\,,\,\,Cx=qxC\,\,,\,\,Cp=qpC.$$
To prove that $D$ is a derivation,
let $a$, $b$ and $c$ be in Z. Define $D(x^ap^bm^c)$ as follows
$$D(x^ap^bm^c)={\sum_{i+j=a-1}x^iAx^jp^bm^c}+
{\sum_{i+j=b-1}x^ap^iBp^jm^c}+{\sum_{i+j=c-1}x^ap^bm^iCm^j}.$$
Then we extend $D$ to all of ${\cal M}_q$ by linearity and continuity.
Now we are going to prove that
$D$ is a derivation. Clearly for $r$, $s$ and $t$ in Z we have
$$D(x^ap^bm^cx^rp^sm^t)=q^{r(b+c)+sc}
D(x^{a+r}p^{b+s}m^{c+t})=$$ $$q^{r(b+c)+sc}
[{\sum_{i+j=a+r-1}x^iAx^jp^{b+s}m^{c+t}}+
{\sum_{i+j=b+s-1}x^{a+r}p^iBp^jm^{c+t}}
+{\sum_{i+j=c+t-1}x^{a+r}p^{b+s}m^iCm^j}].$$
$$=q^{r(b+c)+sc}[{\sum_{i+j=r-1}x^mx^iAx^jp^{b+s}m^{c+t}}+
{\sum_{i+j=a-1}x^iAx^jx^rp^{b+s}m^{c+t}}$$
$$+{\sum_{i+j=b-1}x^ax^rp^iBp^jp^sm^{c+t}}+
{\sum_{i+j=s-1}x^ax^rp^bp^iBp^jm^{c+t}}$$
$$+{\sum_{i+j=c-1}x^ax^rp^bp^sm^iCm^jm^t}+
{\sum_{i+j=t-1}x^ax^rp^bp^sm^cm^iCm^j}]$$
$$={x^ap^bm^c({\sum_{i+j=r-1}x^iAx^jp^sm^t}+
{\sum_{i+j=s-1}x^rp^iBp^jm^t}+{\sum_{i+j=t-1}x^rp^sm^iCm^j})}$$
$$+({\sum_{i+j=a-1}x^iAx^jp^bm^c}+{\sum_{i+j=b-1}x^ap^iBp^jm^c}+
{\sum_{i+j=c-1}x^ap^bm^iCm^j})x^rp^sm^t$$
$$={x^ap^bm^c{D(x^rp^sm^t)}}+{D(x^ap^bm^c)}{x^rp^sm^t}.$$ Therefore,
$D$ is a derivation.

{\bf Corollary 1}. Let $A$, $B$ and $C$ be three elements of
${\cal M}_q$ satisfying the relations of Proposition 1.1.
Then the linear operators $$D_1,D_2,D_3:{\cal M}_q \rightarrow
{\cal M}_q$$ given by
$$D_1(x^ap^bm^c)=\sum_{i+j=a-1}x^iAx^jp^bm^c \;\;,\;\; D_2(p)=
\sum_{i+j=b-1}x^ap^iBp^jm^c \;\;,\;\; D_3(m)=\sum_{i+j=c-1}x^ap^bm^iCm^j,$$
and $$D_1(p)=D_1(m)=0\;\;,\;\;D_2(x)=D_2(m)=0\;\;,\;\;D_3(x)=D_3(p)=0,$$
are derivations.
We call $D_1$ an $x$-derivation, $D_2$ a $p$-derivation and $D_3$ an
$m$-derivation.

{\bf Corollary 2}. Each derivation $D:{\cal M}_q \rightarrow {\cal M}_q$
can be written uniquely
as the sum of an $x$-derivation $D_1$, a $p$-derivation $D_2$,
and an $m$-derivation $D_3$ . Moreover
$$D_1(x)=D(x) \hspace{7.5mm},\hspace{7.5mm} D_2(p)=D(p) \hspace{7.5mm},
\hspace{7.5mm} D_3(m)=D(m).$$

{}From the above we have the following

{\bf Proposition 1.2.} A linear operator $D:{\cal M}_q\rightarrow
{\cal M}_q$ is a derivation
if and only if ${D(x)}$, ${D(p)}$ and ${D(m)}$ are of the following forms
$$D(x)=\overline{\sum}_{k\gg-\infty}{a_k(q)}x^{k+1}p^{-k}m^k
\hspace{7.5mm},\hspace{7.5mm}
D(p)=\overline{\sum}_{k\gg-\infty}{b_k(q)}x^kp^{1-k}m^k,$$
$$D(m)=\overline{\sum}_{k\gg-\infty}{c_k(q)}x^kp^{-k}m^{k+1}.$$
{\bf Corollary}.
A necessary and sufficient condition for the linear operator
$D:{\cal A}_q\rightarrow {\cal M}_q$ to be a derivation,
is that $D(x)$, $D(p)$ and $D(m)$ be of the following forms
$$D(x)=a_1(q)x+a_2(q)pm^{-1}\hspace{5mm},\hspace{5mm} D(p)=b_1(q)p+b_2(q)xm
\hspace{5mm},\hspace{5mm} D(m)=c_1(q)m+c_2(q)x^{-1}p.$$
Now it is clear that for each derivation $D:{\cal M}_q\rightarrow
{\cal M}_q$,
$$D(x)x=xD(x)\hspace{7.5mm},\hspace{7.5mm}
D(p)p=pD(p) \hspace{7.5mm},\hspace{7.5mm} D(m)m=mD(m).$$
More generally

{\bf Lemma 1.3.} Let $D:{\cal M}_q\rightarrow {\cal M}_q$ be a derivation.
Then for each
monomial $z=x^ap^bm^c $ we have $zD(z)=D(z)z$.

{\bf Proof}. Let $z=x^ap^bm^c$. Then
$$zD(z)=x^ap^bm^c[ax^{a-1}D(x)p^bm^c+bx^aD(p)p^{b-1}m^c+cx^ap^bD(m)
m^{c-1}]$$
$$=ax^{a-1}D(x)p^bm^cx^ap^bm^c+bx^aD(p)p^{b-1}m^cx^ap^bm^c+
cx^ap^bD(m)m^{c-1}x^ap^bm^c=D(z)z.$$
Let
$${\frac{\partial}{\partial x}},{\frac{\partial}{\partial p}},
{\frac{\partial}{\partial m}}:{\cal M}_q\rightarrow {\cal M}_q$$
be linear operators given by
$${\frac{\partial}{\partial x}}(x^ip^jm^k)=ix^{i-1}p^jm^k
\hspace{7.5mm},\hspace{7.5mm}
{\frac{\partial}{\partial p}(x^ip^jm^k)=q^{-i}jx^ip^{j-1}m^k}$$.
and $${\frac{\partial (x^ip^jm^k)}{\partial m}}=q^{-(i+j)}kx^ip^jm^{k-1}.$$
{\bf Lemma 1.4.} Assume that $D:{\cal M}_q \rightarrow {\cal M}_q$ is a
derivation. Then $D$
can be written uniquely as $$D=D(x){\frac{\partial}{\partial x}}+
D(p){\frac{\partial}{\partial p}}+D(m){\frac{\partial}{\partial m}}.$$
{\bf Proof.} Let $a,b$ and $c$ be in N. Then
$$D(x^ap^bm^c)=x^aD(p^b)m^c+D(x^a)p^bm^c+x^ap^bD(m^c)$$
$$=bx^ap^{b-1}D(p)m^c+ax^{a-1}D(x)p^bm^c+cx^ap^bm^{c-1}D(m)$$
$$=q^{-a}[bD(p)x^ap^{b-1}m^c]+ax^{b-1}D(x)p^bm^c+
q^{-(i+j)}cD(m)x^ap^bm^{c-1}$$
$$=D(p){\frac{\partial}{\partial p}}+
D(x){\frac{\partial}{\partial x}}+
D(m){\frac{\partial}{\partial m}}](x^ap^bm^c).$$
{\bf Lemma 1.5.} Under the above assumption
$$D_1=D(x){\frac{\partial}{\partial x}}\;\;,\;\;D_2=
D(p){\frac{\partial}{\partial p}}\;\;,\;\;D(m)
{\frac{\partial}{\partial m}}.$$
{\bf Proof.} It is only sufficient to prove that
$D(x){\frac{\partial}{\partial x}}$ ,
$D(p){\frac{\partial}{\partial p}}$ and $D(m){\frac{\partial}{\partial m}}$
are derivations. Let $a,b,c,r,s$ and $t$ be in N. Then
$$D(x){\frac{\partial}{\partial x}}(x^ap^bm^cx^rp^sm^t)
=q^{r(b+c)+sc}D(x){\frac{\partial}{\partial x}}(x^{a+r}p^{b+s}m^{c+t})$$
$$=q^{r(b+c)+sc}
(a+r)D(x)x^{a+r-1}p^{b+s}m^{c+t}$$
$$=x^ap^bm^cD(x){\frac{\partial}{\partial x}}(x^rp^sm^t)+D(x)
{\frac{\partial}{\partial x}}(x^ap^bm^c)x^rp^sm^t.$$
In the same way we see that $D(p){\frac{\partial}{\partial p}}$ and
$D(m){\frac{\partial}{\partial m}}$
are derivations.

The set of all derivations of ${\cal M}_q$ which is clearly a Lie algebra
will be denoted by $\overline \chi({\cal M}_q)$.

{\bf 2 Hamiltonian systems on the quantum plane}

In this section we endow ${\cal M}_q$ with the canonical $q$-deformed
Poisson structure
$$\pi=q^{-\frac 12}{\frac{\partial}{\partial x}}{\wedge}
{\frac{\partial}{\partial p}}
-q^{\frac 12}{\frac{\partial}{\partial p}}{\wedge}{\frac{\partial}
{\partial x}}.$$
The associated $q$-deformed Poisson bracket will be denoted by $\{\;,\;\}_q$.
More precisely,
for each two elements $f,g\in {\cal M}_q$ we have
$$\{f\,,g\}_q=q^{-1/2}{\frac{\partial f}{\partial x}}
{\frac{\partial g}{\partial p}}-q^{1/2}{\frac{\partial f}
{\partial p}}{\frac{\partial g}{\partial x}}.$$

An element $z\in {\cal M}_q $ is called Hamiltonian if the mapping
$$X_z:{\cal M}_q\rightarrow {\cal M}_q$$ defind by $$X_z(f)=\{z\;,\,f\}_q$$
is a derivation. In this case
$X_z$ is called a Hamiltonian derivation.

{\bf Lemma 2.1.} A necessary and sufficient condition for
$z\in {\cal M}_q$ to be Hamiltonian
is that it has the following form
$$z=\overline{\sum}_{k\gg-\infty}{a_k(q)}
x^{k+1}p^{1-k}m^k,\;\;\;a_k(q)\in {\cal M}_q.$$

{\bf Proof}. Let $z$ be of the above form. Then
$$A=\{z\,,x\}=-\overline{\sum}_{k\gg-\infty}{a_k(q)}
q^{-\frac{(1+2k)}{2}}(1-k)x^{1+k}p^{-k}m^k,$$
$$B=\{z\,,p\}_q=\overline{\sum}_{k\gg-\infty}{a_k(q)}q^{-1/2}
(1+k)x^kp^{1-k}m^k \hspace{.5mm},\hspace{.5mm} C={\{z\,,m\}_q}=0.$$
By Proposition 1.2 $X_z$ is a derivation.
Now assume that $X_z$ is a derivation.
Suppose that $z=\sum_{i,j,k\gg-\infty}{a_{ijk}(q)}x^ip^jm^k.$ Then
$X_z(x)=-\sum_{i,j,k}j{a_{ijk}(q)}q^{{1/2}-i}x^ip^{j-1}m^k$ ,
$X_z(p)=\sum_{i,j,k}i{a_{ijk}(q)}q^{-1/2}x^{i-1}p^jm^k$, and $X_z(m)=0.$
But since $X_z$ is a derivation Proposition 1.2 implies that
$i=k+1$ and $j=1-k$. The proof is complete.

The set of all Hamiltonian elements of ${\cal M}_q$ will be denoted by
$H({\cal M}_q)$.
It is clear that $H({\cal M}_q)$ is an $A_0(q)$-module.
Let $z_1=x^{k+1}p^{1-k}m^k$ and
$z_2=x^{l+1}p^{1-l}m^l$. Direct computation shows that
$$\{z_1\,,z_2\}_q=2(k-l)q^{-(1/2+kl)}x^{1+(k+l)}p^{{1-(k+l)}}m^{k+l}.$$
Therefore for each two elements $z_1$ $z_2$ in
$H({\cal M}_q)$, $\{z_1\,,z_2\}_q\in H({\cal M}_q).$
Moreover $\{z_1\,,z_2\}_q=-\{z_2\,,z_1\}_q$. Now let $z_1$ and $z_2$
be as above and
$z_3=x^{n+1}p^{1-n}m^n.$ Then
$$\{\{z_1\,,z_2\}_q\,,z_3\}_q+\{\{z_2\,,z_3\}_q\,,z_1\}_q
+\{\{z_3\,,z_1\}_q\,,z_2\}_q=0.$$
Therefore $(H({\cal M}_q)\;\;\{\,,\}_q)$ is an $A_0(q)$-Lie algebra,
with centre $A_0(q)$. Let
$z_1$ and $z_2$ be as above.Then
$$X_{\{z_1\,,z_2\}_q}(x)=2(k-l)(1-k-l)q^{-(1+k+l+kl)}x^{1+(k+l)}
p^{1-(k+l)}m^{k+l}=[X_{z_1}\;X_{z_2}](x),$$
$$X_{\{z_1\,,z_2\}_q}(p)=2(k-l)(1+k+l)
q^{-(1+kl)}x^{k+l}p^{1-(k+l)}m^{k+l}=[X_{z_1}\,,X_{z_2}](p)$$
and $$X_{\{z_1\,,z_2\}_q}(m)=0.$$
{}From the above considerations we see that the mapping
$$X:H({\cal M}_q)\rightarrow \overline \chi({\cal M}_q)$$
given by $X(z)=X_z$ is a homomorphism of $A_0(q)$-Lie algebras
with kernel $A_0(q)$. Let $z_n\in H({\cal M}_q)$ be defind as follows
$$z_n={1/2}q^{\frac{1-n^2}{2}}x^{1+n}p^{1-n}m^n,\;\;\;\;n\in Z.$$
Then $\{z_m\,,z_n\}=(m-n)z_{m+n}.$
Therefore the Lie algebra of Hamiltonian derivations
is the Virasoro algebra with central charge zero.

The set of all Hamiltonian derivations of ${\cal A}_q$ with values in
${\cal M}_q$
is an $A_1(q)$-Lie algebra generated by
$H=x{\frac{\partial}{\partial x}}-
p{\frac{\partial}{\partial p}}$, $E_+=xm{\frac{\partial}{\partial p}}$,
$E_-=pm^{-1}{\frac{\partial}{\partial x}}$,
with the multiplication rules
$$[E_{+}\,\,E_{-}]=H\;\;,\,\,[H\,\,E_{+}]=
2E_{+}\;\;,\,\,[H\,\,E_{-}]=-2E_{-}.$$
Therefore  this Lie-algebra is $sl(2,A_1(q))$.

{\bf Lemma 2.2.} Let $z$ be in $H({\cal M}_q)$. Then for each
$f$ in ${\cal M}_q$ we have
$$\{z\,,f\}_q=\{z\,,x\}_q{\frac{\partial f}
{\partial x}}+\{z\,,p\}_q{\frac{\partial f}{\partial p}}.$$

{\bf Proof}.
$$\{z\,,f\}_q=X_z(f)=X_z(x){\frac{\partial f}
{\partial x}}+X_z(p){\frac{\partial f}{\partial p}}=
\{z\,,x\}_q{\frac{\partial f}{\partial x}}+
\{z\,,p\}_q{\frac{\partial f}{\partial p}}.$$

By a Hamiltonian system on the quantum plane we mean the triple
$({\cal M}_q,\pi,z)$, where $\pi$ is
the canonical $q$-deformed Poisson structure on ${\cal M}_q$ and
$z\in H({\cal M}_q)$. Let
${\phi}_t$ be a one-parameter group of automorphisms of the
$q$-deformed Poisson
structure $({\cal M}_q,\pi)$. We say that ${\phi}_t$ defines the motion of
the system
$({\cal M}_q,\pi,z)$, if for each $f\in {\cal M}_q$
$${\frac{df_t}{dt}}=\{z_t\,,f_t\}_q,$$
where for each $f\in {\cal M}_q,$ ${\phi}_t(f)$ is denoted by $f_t.$

{\bf Proposition 2.3.} A necessary and sufficient condition for ${\phi}_t$
to define the motion of the Hamiltonian system $({\cal M}_q,\pi,z)$
is that for each $t,$
$z_t\in H({\cal M}_q)$ and $x_t$ and $p_t$ satisfy the following equations
$${\frac{dx_t}{dt}}=\{z_t\,,x_t\}_q\;\;\;\;
\;{\frac{dp_t}{dt}}=\{z_t\,,p_t\}_q \,\,\,\,\,
{\frac{dm_t}{dt}}={\{z_t\,,m_t\}_q}.$$

{\bf Proof}. The condition is clearly necessary. To
prove that the condition is sufficient assume that
$x_t$, $p_t$ and $m_t$ satisfy the above equations and
$z_t\in H({\cal M}_q).$ Then
$$x_t{\frac{dx_t}{dt}}=x_t{\{z_t\,,x_t\}_q}\;\;\;\;p_t
{\frac{dp_t}{dt}}=p_t{\{z_t\,,p_t\}_q}
\;\;\;\;m_t{\frac{dm_t}{dt}}=m_t{\{z_t\,,m_t\}_q}.$$
As we have seen earlier for each $t$, $X_{z_t}$ is a derivation.
Therefore $$x_t{\{z_t\,,x_t\}_q}=x_tX_{z_t}(x_t)
=X_{z_t}(x_t)x_t={\frac{dx_t}{dt}}x_t$$
$${p_t}{\{z_t\,,p_t\}_q}={p_t}X_{z_t}(p_t)=X_{z_t}(p_t){p_t}=
{\frac{dp_t}{dt}}{p_t}.$$
$$m_t{\{z_t\,,m_t\}_q}=m_tX_{z_t}(m_t)=X_{z_t}(m_t)m_t=
{\frac{dm_t}{dt}}m_t.$$
Now let $f=x^ip^jm^k.$ Then $f_t={x_t}^i{p_t}^j{m_t}^k$ and
$${\frac{d{f_t}}{dt}}=i{x_t}^{i-1}{\frac{d{x_t}}{dt}}{p_t}^j{m_t}^k+
{x_t}^ij{p_t}^{j-1}{\frac{d{p_t}}{dt}}{m_t}^k+
{x_t}^i{p_t}^jk{m_t}^{k-1}{\frac{dm_t}{dt}}$$
$$={\frac{d{x_t}}{dt}}(i{x_t}^{i-1}{p_t}^jm^k)+
{\frac{d{p_t}}{dt}}(q^{-i}j{x_t}^i{p_t}^{j-1}m^k)
+{\frac{dm_t}{dt}}(q^{-(i+j)}k{x_t}^i{p_t}^j{m_t}^{k-1})$$
$$={X_{z_t}(x_t)}{\frac{\partial{f_t}}
{\partial{x_t}}}+{X_{z_t}(p_t)}{\frac{\partial{f_t}}
{\partial{p_t}}}+X_{z_t}(m_t){\frac{\partial f_t}{\partial x_t}}
=X_{z_t}(f_t)={\{z_t\,,f_t\}_q}.$$
Therefore for each $f\in{\cal M}_q$ we have ${\frac{df_t}{dt}}=
{\{z_t\,,f_t\}_q}.$

Now since for each $t$, ${\{z_t\,,z_t\}_q}=0$ therefore
${\frac{d{z_t}}{dt}}=0.$
This means that $z$ is an invariant of motion.
It is easy to see that any analytic function of
$z$ is also an invariant of motion.

{\bf Proposition 2.4.} Let $z\in H({\cal M}_q).$ Then the Cauchy problem
$${\frac{d{x_t}}{dt}}={\{z\,,x_t\}_q}
\;\;,\;\;{\frac{d{p_t}}{dt}}={\{z\,,p_t\}_q},
\;\;,\;\;{\frac{dm_t}{dt}}={\{z\,,m_t\}_q}\;,\;x_0=x\;,\;p_0=p\;,\;m_0=m$$
has a unique generalized $q$-meromorphic solution. Moreover the solution
satisfies the relations
${p_t}{x_t}=q{x_t}{p_t}$ and $m_t=m.$

{\bf Proof}. Let $D=X_z$. Then the above Cauchy problem has
the unique solution
$$x_t={e^{tD}}x\;\;,\;\;\;p_t={e^{tD}}p\;\;,\;\;m_t={e^{tD}}m.$$
It is easy to see that for each $t$, $x_t$ and $p_t$ are generalized
$q$-meromorphic functions.
Let $z=\overline{\sum}_{k\gg-\infty}{a_k(q)}x^{1+k}p^{1-k}m^k.$ Then
$$D=(- \overline{\sum}_{k\gg-\infty}{a_k(q)}
q^{-(\frac{1+k}{2}}(1-k)x^{1+k}p^{-k}m^k){\frac{\partial}{\partial x}}
+(\overline{\sum}_{k\gg-\infty}{a_k(q)}
q^{-1/2}(1+k)x^kp^{1-k}m^k){\frac{\partial}{\partial p}}.$$
Now from the relations
$$x_t={e^{tD}}x=\sum_{n=0}^{\infty}
{\frac{t^nD^n}{n!}}x\;\;,\;\;p_t={e^{tD}}p=\sum_{k=0}^{\infty}
{\frac{t^kD^k}{k!}}p\;\;,\;\;
m_t={e^{tD}}m=\sum_{l=0}^{\infty}{\frac{t^lD^l}{n!}}m.$$
and using the above expression for $D$ we see that $p_tx_t=qx_tp_t$, and
$m_t=m.$

Notice that the Hamilton equations on ${\cal M}_q$ in general
does not define a motion of the
corresponding Hamiltonian system, in our sense.
In the following we prove that the situation for ${\cal A}_q$ is different.
Consider the Hamiltonian system $({\cal M}_q\,,\pi\,,z)$, where
$z={\alpha}{\frac{p^2m^{-1}}2}+{\beta}{\frac{x^2m}2}+{\gamma}xp,$ where
${\alpha}\,,{\beta}$ and ${\gamma}\in A_1(q).$ The corresponding Hamilton
equations are
$${\frac{dx_t}{dt}}=-q^{-1/2}({\alpha}p_tm^{-1}+q^{-1}
{\gamma}x_t)\,\,\,,\,\,\,
{\frac{dp_t}{dt}}=q^{-1/2}({\beta}x_tm+{\gamma}p_t).$$
Or in the matrix form
$$(\frac{dx_t}{dt}\;\;\;\; \frac{dp_t}{dt})=(x_t\;\;\; p_t)
\left( \begin{array}{cc}
-q^{-1/2}\gamma & q^{-1/2}\beta m\\
-q^{1/2}\alpha m^{-1} & q^{-1/2}\gamma \end{array}
\right).$$
By solving this linear differential equation with constant
coefficients we obtain

$$\left\{ \begin{array}{lr}
x_t={\cosh{\theta}t}x-{\theta}^{-1}{\sinh{\theta}t}(q^{-1/2}
{\gamma}x+q^{1/2}{\alpha}pm^{-1})\\
p_t={\cosh{\theta}t}p+{\theta}^{-1}{\sinh{\theta}t}
(q^{-1/2}{\beta}xm+q^{-1/2}{\gamma}p)
\end{array} \right.$$
where ${\theta}=(q^{-1/2}{\gamma}^2-{\alpha}{\beta})^{1/2}
\,\,\,,\,\,\,x=x_0\,\,\,,\,\,\,p=p_0.$

Now let $q$ be a constant complex number and let $z^{1/2}$
denotes the non-principal branch of
the second root of $z.$ Then

1. Let $\alpha=1$ and $\beta=\gamma=0.$ In this case we have
$$x_t=x-q^{1/2}pm^{-1}\;\;\;,\;\;\;p_t=p.$$
2. Let $\alpha=1\;\; \beta=\omega^2$ and $\gamma=0$.In this case we have
$$x_t=x{\cos{\omega}t}-q^{-1/2}p{({\omega}m)}^{-1}{\sin{\omega}t}$$
$$p_t=p{\cos{\omega}t}+q^{-1/2}{\omega}xm{\sin{\omega}t}.$$

Notice that in these two special cases the slight difference between our
results and those
in [5] comes from the difference between the definitions of the q-deformed
Poisson structures
given in [5] and in this paper and the difference
between Hamiltonians in [5] and here
come from different rules of differentiation.


\begin{thebibliography}{99}
\bibitem{1}R. Abraham and E. Marsden, {\it Foundationsd of Mechanics}, 1978.
\bibitem{2}V. Arnold, {\it Mathematical Methods of Classical Mechanics},Mir
Publ. (Moscow 1975).
\bibitem{3}A. Shafei Deh Abad and V. Milani, JMP 35(1994)5074.
\bibitem{4}S. V. Shabanov, J.Phys.A:Math.Gen.26(1993)2583.
\bibitem{5}I. Ya. Arefeva and I. V. Volovich, Phys.letter B,268(1991)179.
\bibitem{6}J. Schwenk and J. Wess, Phys.letter B,291(1992)273.
\end{thebibliography}
\end{document}